\documentclass[11pt,a4paper]{article}
\usepackage{amsmath,amssymb}
\usepackage{epsfig,graphicx}

\begin{document}

\title{\bf Effective action and Schwinger-DeWitt technique
in DGP brane models}
\author{
A.~O.~Barvinsky$^{a}$\footnote{{\bf e-mail}: barvin@lpi.ru}\,,\;
D.~V.~Nesterov$^a$\footnote{{\bf e-mail}: nesterov@lpi.ru}
\\
$^a$ \small{\em Theory Department, Lebedev Physics Institute,} \\
\small{\em Leninsky prospect 53, Moscow, Russia, 119991.}
}
\date{}
\maketitle

\begin{abstract}
We give blueprints of the Schwinger-DeWitt technique for the
covariant curvature expansion of quantum effective action for the
DGP type models in curved spacetime.
\end{abstract}

\section{Introduction}
  \hspace{\parindent}
Modified theories of gravity in the form of braneworld models can in
principle account for the phenomenon of dark energy as well as for
nontrivial compactifications of multi-dimensional string models. It
becomes increasingly more obvious that one should include in such
models the analysis of quantum effects beyond the tree-level
approximation \cite{quantumDGP}. This is the only way to reach an
ultimate conclusion on the resolution of such problems as the
presence of ghosts \cite{ghosts} and low strong-coupling scale
\cite{scale}. Quantum effects in brane models are also important for
the stabilization of extra dimensions \cite{GarPujTan}, fixing the
cross-over scale in the Brans-Dicke modification of the DGP model
\cite{Pujolas} and in the recently suggested mechanism of the
cosmological acceleration generated by the four-dimensional
conformal anomaly \cite{slih}.

A general framework for treating quantum effective actions in brane
models (or, more generally, models with timelike and spacelike
boundaries) was recently suggested in
\cite{BKRK,gospel,qeastb,toyDGP}. The main peculiarity of these
models is that due to quantum field fluctuations on the branes the
field propagator is subject to generalized Neumann boundary
conditions involving normal and tangential derivatives on the
brane/boundary surfaces. This presents both technical and conceptual
difficulties, because such boundary conditions are much harder to
handle than the simple Dirichlet ones. The method of \cite{qeastb}
provides a systematic reduction of the generalized Neumann boundary
conditions to Dirichlet conditions. As a byproduct it disentangles
from the quantum effective action the contribution of the surface
modes mediating the brane-to-brane propagation, which play a very
important role in the zero-mode localization mechanism of the
Randall-Sundrum type \cite{RS}. The purpose of this work is to make
the next step --- to extend a well-known Schwinger-DeWitt technique
\cite{DeWitt,PhysRep,McKean-Singer,Vassilevich} to the calculation
of this contribution in the DGP model in a weakly curved spacetime
in the form of the {\em covariant} curvature expansion.

Briefly the method of \cite{qeastb} looks as follows. The action of
a (free field) brane model generally contains the bulk and the brane
parts,
    \begin{eqnarray}
    S[\,\phi\,]=\frac12\int_{\rm\bf B} d^{d+1}X\,G^{1/2}\phi(X)
    F(\nabla_X)\,\phi(X)
    +\frac12\int_{\rm \bf b}
    d^dx \,g^{1/2}\varphi(x)\,
    \kappa(\nabla_x)\,\varphi(x) \ ,                         \label{1}
    \end{eqnarray}
where the $(d\!+\!1)$-dimensional bulk and the $d$-dimensional brane
coordinates are labeled respectively by $X=X^A$ and $x=x^\mu$, and
the boundary values of bulk fields $\phi(X)$ on the brane/boundary
${\rm\bf b}=\partial\rm\bf B$ are denoted by $\varphi(x)$,
    \begin{eqnarray}
    \phi(X)\,\Big|_{\,\rm\bf b}=\varphi(x),        \label{2}
    \end{eqnarray}
$G$ and $g$ are the determinants of the bulk $G_{AB}$ and
$g_{\mu\nu}$ metrics respectively.

The kernel of the bulk Lagrangian is given by the second order
differential operator $F(\nabla_X)$, whose covariant derivatives
$\nabla_X$ are integrated by parts in such a way that they form
bilinear combinations of first order derivatives acting on two
different fields. Integration by parts in the bulk gives nontrivial
surface terms on the brane/boundary. In particular, this operation
results in the Wronskian relation for generic test functions
$\phi_{1,\,2}(X)$,
    \begin{eqnarray}
    \int_{\rm\bf B} d^{\,d+1}X\,G^{1/2}
    \left(\,\phi_1\stackrel{\rightarrow}{F}\!(\nabla_X)\phi_2-
    \phi_1\!\stackrel{\leftarrow}{F}\!(\nabla_X)\,\phi_2\right)=
    -\int_{\partial{\rm\bf B}} d^{\,d}x\,g^{1/2}
    \left(\,\phi_1\stackrel{\rightarrow}{W}\!
    \phi_2-
    \phi_1\stackrel{\leftarrow}{W}\!
    \phi_2\right).                          \label{3a}
    \end{eqnarray}         
Arrows everywhere here indicate the direction of action of
derivatives either on $\phi_1$ or $\phi_2$.

The brane part of the action contains as a kernel some local
operator $\kappa(\nabla)$, $\nabla\equiv\nabla_x$. Its order in
derivatives depends on the model in question. In the Randall-Sundrum
model \cite{RS}, for example, it is for certain gauges just an
ultralocal multiplication operator generated by the tension term on
the brane. In the Dvali-Gabadadze-Porrati (DGP) model \cite{DGP}
this is a second order operator induced by the brane Einstein term
on the brane, $\kappa(\nabla)\sim\nabla\nabla/m$, where $m$ is the
DGP scale which is of the order of magnitude of the horizon
scale, being responsible for the cosmological acceleration
\cite{Deffayet}. In the context of the Born-Infeld action in D-brane
string theory with vector gauge fields, $\kappa(\nabla)$ is a
first-order operator \cite{open}.

In all these cases the variational procedure for the action
(\ref{1}) with dynamical (not fixed) fields on the boundary
$\varphi(x)$ naturally leads to generalized Neumann boundary
conditions of the form
    \begin{eqnarray}
    \left.\Big(\stackrel{\rightarrow}{W}\!(\nabla_X)
    +\kappa(\nabla)\Big)\,\phi\,\right|_{\,\rm\bf b}
    =0,                                                     \label{3}
    \end{eqnarray}
which uniquely specify the propagator of quantum fields and,
therefore, a complete Feynman diagrammatic technique for the system
in question. The method of \cite{qeastb} allows one to
systematically reduce this diagrammatic technique to the one subject
to the Dirichlet boundary conditions $\phi|_{\,\rm\bf b}=0$. The
main additional ingredient of this reduction procedure is the brane
operator $\mbox{\boldmath$F$}^{\,\rm brane}(x,x')$ which is
constructed from the Dirichlet Green's function $G_D(X,X')$ of the
operator $F(\nabla)$ in the bulk,
    \begin{eqnarray}
    \mbox{\boldmath$F$}^{\,\rm brane}(x,x')=-
    \stackrel{\rightarrow}{W}\!(\nabla_X\!)\,G_{D}(X,X')\!
    \stackrel{\leftarrow}{W}\!(\nabla_{X'}\!)
    \,\Big|_{\,X=e(x),\,X'=e(x')}
    +\kappa(\nabla)\,\delta(x,x')\ .           \label{5}
    \end{eqnarray}
This expression expresses the fact that the kernel of the Dirichlet
Green's function is being acted upon both arguments by the Wronskian
operators with a subsequent restriction to the brane, with $X=e(x)$
denoting the brane embedding function.

As shown in \cite{qeastb}, this operator determines the
brane-to-brane propagation of the physical modes in the system with
the classical action (\ref{1}) (its inverse is the brane-to-brane
propagator) and additively contributes to its full one-loop
effective action according to
    \begin{eqnarray}
    \mbox{\boldmath$\varGamma$}_{\rm 1-loop}\equiv\frac12\;
    {\rm Tr}_N^{(d+1)}\ln F=\frac12\;{\rm Tr}_D^{(d+1)}\ln F
    +\frac12\;{\rm Tr}^{(d)}
    \ln \mbox{\boldmath$F$}^{\,\rm brane},    \label{11}
    \end{eqnarray}
where ${\rm Tr}_{D,\,N}^{(d+1)}$ denotes functional traces of the
bulk theory subject to Dirichlet and Neumann boundary conditions,
respectively, while ${\rm Tr}^{(d)}$ is a functional trace in the
boundary $d$-dimensional theory. The full quantum effective action
of this model is obviously given by the functional determinant of
the operator $F(\nabla_X)$ subject to the generalized Neumann
boundary conditions (\ref{5}), and the above equation reduces its
calculation to that of the Dirichlet boundary conditions plus the
contribution of the brane-to-brane propagation.

Here we apply (\ref{11}) to a simple model of a scalar field which
mimics in particular the properties of the brane-induced gravity
models and the DGP model \cite{DGP}. This is the $(d\!+\!1)$-dimensional
massive scalar field $\phi(X)=\phi(x,y)$ with mass $M$ living in the
{\em curved} half-space $y\geq 0$ with the additional
$d$-dimensional kinetic term for $\varphi(x)\equiv\phi(x,0)$
localized at the brane (boundary) at $y=0$,
    \begin{eqnarray}
    S[\,\phi\,]=\frac12\int\limits_{y\geq 0}
    d^{d+1}X\,G^{1/2}\Big((\nabla_X\phi(X))^2
    +M^2\phi^2(X)\Big)
    +\frac1{4m}\int
    d^dx \,g^{1/2}\,(\nabla_x\varphi(x))^2.           \label{1.1}
    \end{eqnarray}
Here and in what follows we work in a Euclidean
(positive-signature) spacetime. Therefore, this action corresponds
to the following choice of $F(\nabla_X)$ in terms of
$(d\!+\!1)$-dimensional and $d$-dimensional covariant D'Alembertians
(Laplacians)
    \begin{eqnarray}
    &&F(\nabla_X)=M^2-\Box^{(\,d+1)}=
    M^2-G^{AB}\nabla_A\nabla_B,.              \label{6}
    \end{eqnarray}
In the normal Gaussian coordinates its Wronskian operator is given
by $W=-\partial_y$ --- the normal derivative with respect to
outward-pointing normal to the brane,  and the boundary operator
$\kappa(\nabla)$ equals
    \begin{eqnarray}
    \kappa(\nabla)=-\frac1{2m}\,\Box,\, \,\,\,\,
    \Box=\Box^{(d)}\equiv g^{\mu\nu}\nabla_\mu\nabla_\nu, \label{8}
    \end{eqnarray}
where the dimensional parameter $m$ mimics the role of the DGP
scale \cite{DGP}. Thus, the generalized Neumann boundary conditions
in this model involve second-order derivatives tangential to the
brane,
    \begin{eqnarray}
    \Big(\partial_y
    +\frac1{2m}\,\Box\Big)\,\phi\,
    \Big|_{\,\rm\bf b}=0,               \label{1.3}
    \end{eqnarray}
cf. (\ref{3}) with $W=-\partial_y$ and $\kappa=-\Box/2m$.

As was shown \cite{toyDGP}, the flat space brane-to-brane operator for
such a model has the form of the pseudodifferential operator with
the flat-space $\Box$,
    \begin{eqnarray}
    \mbox{\boldmath$F$}^{\,\rm brane}(\nabla)
    =\frac1{2m}\,(-\Box+2m\sqrt{M^2-\Box}).          \label{9}
    \end{eqnarray}
In the massless case of the DGP model \cite{DGP}, $M=0$, this
operator is known to mediate the gravitational interaction on the
brane, interpolating between the four-dimensional Newtonian law at
intermediate distances and the five-dimensional law at the horizon
scale $\sim 1/m$ \cite{scale}.

Here we generalize this construction to a curved spacetime and
expand the brane-to-brane operator and its effective action in
covariant curvature series. This is the expansion in powers of the
bulk curvature symbolically denoted below as $R$, extrinsic
curvature of the brane denoted by $K$ and their covariant
derivatives --- all taken at the location of the brane. The
expansion starts with the approximation (\ref{9}) based on the {\em
full covariant} d'Alembertian on the brane. We present a systematic
technique of calculating curvature corrections to (\ref{9}) and
rewrite their nonlocal operator coefficients --- functions of the
covariant $\Box$
--- in the form of the generalized (weighted) proper time
representation.

The success of the conventional Schwinger-DeWitt method is based on
the fact that the one-loop effective action of the operator, say
$M^2-\Box$, has a proper time representation
    \begin{eqnarray}
    &&\frac12\,{\rm Tr}\,\ln\,\Big(M^2-\Box\Big)
    =-\frac12 \int_0^\infty
    \frac{ds}s\;e^{-s\,M^2}\,
    {\rm Tr}\:e^{s\,\Box}.               \label{proptime}
    \end{eqnarray}
In view of the well-known small time expansion for the heat kernel
\cite{DeWitt,PhysRep},
    \begin{eqnarray}
    e^{\,s\,\Box}\delta(x,x')=\frac1{(4\pi s)^{d/2}}
    D^{1/2}(x,x')\,e^{-\sigma(x,x')/2s}\sum_{n=0}^\infty\,s^n\,
    a_n(x,x'),
    \end{eqnarray}
($\sigma(x,x')$ is the geodetic world function, $D(x,x')$ is the
associated Van Vleck determinant and $a_n(x,x')$ are the
Schwinger-DeWitt or Gilkey-Seely coefficients) the curvature
expansion eventually reduces to the calculation of the coincidence
limits of $a_n(x,x')$ and a trivial proper time integration
resulting in the inverse mass expansion
    \begin{eqnarray}
    \frac12\,{\rm Tr}\,\ln\,\Big(M^2-\Box\Big)
    =-\frac12 \frac1{(4\pi)^{d/2}}
    \sum_{n=0}^\infty\frac{\Gamma(n\!-\!d/2)}{M^{2n-d}}\,\int
    dx\,g^{1/2}\,a_n(x,x).
    \end{eqnarray}

As we will show below, the calculation of the brane effective action
differs from the conventional Schwinger-DeWitt case in that the
proper time integral (\ref{proptime}) contains in the integrand a
certain extra weight function $w(s)$ and instead of just ${\rm
Tr}\,e^{s\Box}$ one has to calculate the trace of the heat kernel
acted upon by a certain local differential operator ${\rm
Tr}\,\big(W(\nabla)e^{s\Box}\big)$. This again reduces to the
calculation of the coincidence limits --- this time of the multiple
covariant derivatives of $a_n(x,x')$,
$\nabla_{\mu_1}...\nabla_{\mu_n}a_n(x,x')|_{x'=x}$ --- the task
easily doable within a conventional DeWitt recurrence procedure for
$a_n(x,x')$.

\section{Perturbation theory for the bulk Green's function and brane
effective action}
  \hspace{\parindent}
In normal Gaussian coordinates the covariant bulk d'Alembertian
decomposes as $\Box^{\,(d+1)}_X=\partial_y^2+\Box(y)+...$, where
ellipses denote depending on spin terms at most linear in
derivatives\footnote{This term for a general spin structurally has the form
$K\nabla_X+K^2+(\nabla K)+R$ where $K$ is the extrinsic curvature of
$y={\rm const}$ slices and $R$ is the bulk curvature.} and $\Box(y)$
is a covariant d'Alembertian on the slice of constant coordinate
$y$. Therefore the full bulk operator takes the form
    \begin{eqnarray}
    F(\nabla)=M^2-\Box^{(\,d+1)}_X+P(X)=
    M^2-\Box-\partial_y^2-V(X\,|\,\partial_y,\nabla)\equiv F^0-V,
    \end{eqnarray}
in which all nontrivial $y$-dependence is isolated as a perturbation
term $V(X\,|\,\partial_y,\nabla)\equiv V(y,\partial_y)$ --- a
first-order differential operator in $y$, proportional to the
extrinsic and bulk curvatures, and of second order in brane
derivatives $\nabla$ which we do not explicitly indicate here by
assuming that they are encoded in the operator structure of
$V(y,\partial_y)$. In particular, it includes the difference
$\Box(0)-\Box(y)\equiv\Box-\Box(y)$ expandable in Taylor series in
$y$.

The kernel of the bulk Green's function can formally be written as a
$y$-dependent nonlocal operator acting on the $d$-dimensional brane
--- some non-polynomial function of the brane covariant derivative
    \begin{eqnarray}
    G_D(X,X')=G_D(y,y'|\,\nabla)\,\delta(x,x').
    \end{eqnarray}
The perturbation expansion for $G_D(y,y'|\,\nabla)$ is usual
    \begin{eqnarray}
    G_D=G_D^0+G^0_D V G^0_D+...=
    G^0_D\sum_{n=0}^\infty \big( G^0_D\,
    V \big)^n,                           \label{Gpert}
    \end{eqnarray}
where $G_D^0$ is the propagator for operator $F^0$ obeying Dirichlet boundary conditions
and  the composition law includes the integration over the bulk
coordinates, like for example in the first subleading term
    \begin{eqnarray}
    G^0_D V G^0_D(y,y')=\int_0^\infty dy''\,
    G^0_D(y,y'') V(y'',\partial_{y''}) G^0_D(y'',y').
    \end{eqnarray}

The lowest order Green's function in the half-space of the DGP model
setting --- the Green's function of $F^0=M^2-\Box-\partial_y^2$
subject to Dirichlet conditions on the brane $y=0$ and at infinity
--- reads as follows
    \begin{eqnarray}
    &&G_{D}^0(y,y')=\frac{e^{-|\,y-y'|\,\sqrt{M^2-\Box}}
    -e^{-(y+y')\,\sqrt{M^2-\Box}}}{2\sqrt{M^2-\Box}}.
    \end{eqnarray}
We want to stress that here we assume the exact (curved)
$d$-dimensional d'Alembertian $\Box$ depending on the induced metric
of the brane $g_{\mu\nu}(x)$. This means that in the lowest order
approximation the underlying spacetime is not flat, but rather has a
nontrivial but constant in $y$ metric of constant $y$ slices.
Correspondingly in the zeroth order we have
    \begin{eqnarray}
    \big[\stackrel{\rightarrow}{W} G_{D}(y,y')\!
    \stackrel{\leftarrow}{W}
    \big]_{\,y=y'=0}^{\,0}=\,\,
    \stackrel{\rightarrow}{\partial_y}G_{D}^0(y,y')
    \stackrel{\leftarrow}{\partial_y}\!
    \,\Big|_{\,y=y'=0}=-\sqrt{M^2-\Box}. \label{G0}
    \end{eqnarray}

The perturbation of the bulk operator can be expanded in Taylor
series in $y$, so that it reads
    \begin{eqnarray}
    V(y,\partial_y)=
    \sum_{k=0}^\infty y^k\,V_k(\partial_y),       \label{V}
    \end{eqnarray}
where $V_k(\partial_y)=V_k(\partial_y|\nabla)$ is a set of
$y$-independent {\em local $d$-dimensional covariant} operators of
second order in $\nabla_x$ and first order in $\partial_y$.

On substitution of (\ref{G0}) and (\ref{V}) into (\ref{Gpert})
exactly calculable integrals over $y$ result in a nonlocal series in
inverse powers of $\sqrt{M^2-\Box}$, and the perturbation expansion
takes the form
    \begin{eqnarray}
    \big[\stackrel{\rightarrow}{W} G_{D}(y,y')\!
    \stackrel{\leftarrow}{W}
    \big]_{\,y=y'=0}=-\sqrt{M^2-\Box}
    +\sum_{k=0}^\infty U_k(\nabla)\frac1{(M^2-\Box)^{k/2}},
    \end{eqnarray}
where $U_k(\nabla)$ is a set of local covariant differential
operators acting on the brane\footnote{Strictly speaking each $k$-th
order of this series arises in the form of the following nonlocal
chain of square root ``propagators",
$\frac1{(M^2-\Box)^{l_1/2}}\,U_1\frac1{(M^2-\Box)^{l_2/2}}\,U_2...
U_{p-1}\frac1{(M^2-\Box)^{l_p/2}}$, $l_1+l_2+...+l_p=k$, with
differential operators $U_i$ as its vertices, but all these
propagators can be systematically commuted either to the uppermost
right or left by the price of extra commutator terms of the same
structure.}. The dimensionality of each $U_k(\nabla)$ is the inverse
length to the power $k\!+\!1$, which is composed of the dimensionalities
of bulk and extrinsic curvatures and covariant derivatives all taken
on the brane at $y=0$.

With $\kappa(\nabla)=-\Box/2m$ the brane-to-brane operator reads
    \begin{eqnarray}
    2m\mbox{\boldmath$F$}^{\,\rm brane}(\nabla)
    =-\Box+2m\sqrt{M^2-\Box}
    -2m\sum_{k=0}^\infty U_k(\nabla)\frac1{(M^2-\Box)^{k/2}}.
    \end{eqnarray}
Then we consider the perturbation series for the functional trace of
its logarithm in powers of the full $U_k$-series. After reexpansion
in powers of two sets of nonlocal propagators $1/\sqrt{M^2-\Box}$
and $1/(-\Box+2m\sqrt{M^2-\Box})$ the brane effective action finally
takes the form
    \begin{eqnarray}
    &&\frac12\;{\rm Tr}
    \ln \mbox{\boldmath$F$}^{\,\rm brane}=
    \frac12 {\rm Tr}
    \ln\Big(-\Box+2m\sqrt{M^2-\Box}\,\Big)\nonumber\\
    &&\qquad\qquad\qquad
    +\sum_{k\geq 0,\,l\geq 1} {\rm Tr}\; W_{kl}(\nabla)
    \frac1{(M^2-\Box)^{k/2}}\,
    \frac1{(-\Box+2m\sqrt{M^2-\Box}\,)^l}   \label{efacpert}
    \end{eqnarray}
with a new set of local covariant differential operators
$W_{kl}(\nabla)$ acting on the brane. The dimensionality of
$W_{kl}(\nabla)$ is $k+2l$ in units of inverse length. One should
also remember that each power of $1/(-\Box+2m\sqrt{M^2-\Box})$ is
accompanied by one power of $m$ in the numerator, so that
structurally
    \begin{eqnarray}
    W_{kl}(\nabla)\sim m^l\,\nabla^a\,R^b\,K^c,
    \end{eqnarray}
where the integer overall powers of the covariant derivatives
$\nabla$, bulk curvatures $R$ and extrinsic curvatures $K$ are
constrained by the relation $a+2b+c=k+l$.

 \section{Generalized proper time method}
  \hspace{\parindent}
Our goal now is to find the proper time representation of nonlocal
operators in Eq.(\ref{efacpert}) in the form of the exponentiated
$\Box$. A systematic way to do this consists in the following
factorization of the brane-to-brane operator as
    \begin{eqnarray}
    2m\mbox{\boldmath$F$}^{\,\rm brane}_0(\nabla)
    =-\Box+2m\sqrt{M^2-\Box}=
    (\sqrt{M^2-\Box}-m_+)(\sqrt{M^2-\Box}-m_-).   \label{factorization}
    \end{eqnarray}
Here the masses $m_\pm$ are the roots of the relevant quadratic
equation, $x^2+2mx-M^2=0$, $x=\sqrt{M^2-\Box}$,
    \begin{eqnarray}
    m_\pm=-m\pm\sqrt{M^2+m^2},\qquad m_-\!<\!-M<0<m_+\!<M,
    \end{eqnarray}
which determine the poles of the propagator of
$\mbox{\boldmath$F$}^{\,\rm brane}_0(\nabla)$ in spacetime with the
Lorentzian signature\footnote{The pole at $\Box_-$ is formally
tachyonic, but it is located on the unphysical sheet of the Riemann
surface for the propagator in the complex plane of $\Box$
\cite{GabShif} (which is indicated in (\ref{phase}) by the
nontrivial phase). Moreover, its residue is identically vanishing in
view of $m_-<0$. Therefore this pole does not correspond to a real
particle. For $M\neq 0$ only $\Box_+$ gives rise to a particle with
the decreasing mass as $M\to 0$, $M^2-m_+^2\to 0$, which also
disappears in the DGP limit because the pole residue also vanishes
at $M=0$. In this limit only the continuum spectrum of massive
intermediate states survives forming the spectral representation for
the DGP propagator \cite{DHK}
    \begin{eqnarray}
    \frac1{-\Box+2m\sqrt{-\Box}}=\frac{4m}\pi\int_0^\infty
    \frac{d\mu}{\mu^2+4m^2}\,\frac1{\mu^2-\Box}.
    \end{eqnarray}}
    \begin{eqnarray}
    &&\Box_+=M^2-m_+^2>0,        \label{realparticle}\\
    &&\Box_-=|M^2-m_-^2|\,e^{3i\pi}<0.   \label{phase}
    \end{eqnarray}

The factorization (\ref{factorization}) allows one to rewrite the
$l$-th power of the brane-to-brane propagator in (\ref{efacpert}) as
    \begin{eqnarray}
    \frac1{(-\Box+2m\sqrt{M^2-\Box}\,)^l}=
    \frac1{(\sqrt{M^2-\Box}-m_+)^l}\,\frac1{(\sqrt{M^2-\Box}-m_-)^l}
    \end{eqnarray}
and then decompose the resulting fraction into the sum of simple
fractions for which one has explicit proper time representations.
These representations begin with the following relations \cite{Abramowitz_Stegun}
    \begin{eqnarray}
    &&\frac1{(M^2-\Box)^{k/2}}=
    \frac1{\varGamma({k/2})}
    \int\limits_0^\infty ds\;s^{k/2-1}\,
    e^{\,s(\Box-M^2)},                    \label{proptimerep1}\\
    &&\frac1{\sqrt{M^2-\Box}-m}=
    \int\limits_0^\infty ds\;e^{\,s(\Box-M^2)}\,
    \left(\frac1{\sqrt{\pi s}}
    +m\, w(-m\sqrt{s})\right),\,\,\,m<M,  \label{proptimerep2}\\
    &&\frac1{\sqrt{M^2-\Box}\,\big(\sqrt{M^2-\Box}-m\big)}=
    \int\limits_0^\infty
    ds\;e^{\,s(\Box-M^2)}\,
    w(-m\sqrt{s}),\,\,\,m<M,                  \label{proptimerep3}
    \end{eqnarray}
which generate (by differentiating with respect to $m$, $\Box$, $M$
and linear recombining the results) the list of fractions with all
possible powers of the factors $\sqrt{M^2-\Box}$ and
$\sqrt{M^2-\Box}-m_\pm$ in their denominators. Here the weight
function $w(s)$ is given in terms of the error function \cite{Abramowitz_Stegun}
$\mathrm{erf}(x)=\frac2{\sqrt\pi}\int_0^x dy\,\exp(-y^2)$ and has the
following ultraviolet and infrared asymptotics
    \begin{eqnarray}
    w(x)\equiv e^{x^2}
    \Big(1-\mathrm{erf}(x)\Big)
    \to\left\{\begin{array}{ll}\,\,\;\,1\,,\,\,&\,x\to 0,\\
    \,\frac1{x\sqrt\pi}\,,\,\,\,\,\,&\,x
    \to +\infty,\\
    \,2\,e^{x^2},\,\,\,\,\,&\,x
    \to -\infty\end{array}\right.     \label{wasymp}
    \end{eqnarray}

The last two proper time integrals above are defined only for $m<M$
(for any negative $m$ and for $m<M$ if $m$ is positive), because in
view of these asymptotics they are convergent at infinity only in
this range. Interestingly, the forbidden domain corresponds to the
real tachyon, because for $m>M>0$ the pole $\Box=M^2-m^2$ belongs to
the physical sheet of the propagator, and its residue is
nonvanishing.

From (\ref{proptimerep2})-(\ref{proptimerep3}) it immediately
follows that the zeroth order brane-to-brane propagator and its
one-loop functional determinant read
    \begin{eqnarray}
    &&\frac1{-\Box+2m\sqrt{M^2-\Box}}=
    \int\limits_0^\infty ds\;e^{s\,(\Box-M^2)}\;
    \frac{m_+w(-m_+\sqrt{s})-m_-w(-m_-\sqrt{s})}{m_+-m_-},  \label{wpm}\\
    &&{\rm Tr}\,\ln\Big(\!-\Box+2m\sqrt{M^2-\Box}\,\Big)\nonumber\\
    &&\qquad\qquad\qquad\qquad=
    -\frac12\,{\rm Tr}\,
    \int\limits_0^\infty\frac{ds}s\,
    e^{s\,(\Box-M^2)}\,\Big(w(-m_+\sqrt{s})
    +w(-m_-\sqrt{s})\Big),                        \label{zeroefac}
    \end{eqnarray}

For the DGP model case with $M^2=0$ and $m_+=0$, $m_-=-2\,m$ these
representations simplify to the equations derived in \cite{toyDGP}
    \begin{eqnarray}
    &&\frac1{-\Box+2m\sqrt{-\Box}}=
    \int\limits_0^\infty ds\,e^{s\,\Box}\,w(\,2m\sqrt{s}),  \label{7.1}\\
    &&{\rm Tr}\,\ln\Big(\!-\Box+2m\sqrt{-\Box}\,\Big)=
    -{\rm Tr}\,
    \int\limits_0^\infty\frac{ds}s\,
    e^{s\,\Box}\,\frac{1+w(\,2m\sqrt{s})}2.       \label{7.2}
    \end{eqnarray}
The interpretation of the weight contribution here is very
transparent. It interpolates between the ultraviolet and infrared
domains where the brane operator and its logarithm have
qualitatively different behaviors. In the domain of a small proper
time $m\sqrt{s}\ll 1$ it approximates the brane operator by a large
$-\Box\gg m^2$ (hence the overall weight $(1+w)/2\simeq 1$), whereas
in the infrared domain $m\sqrt{s}\gg 1$ it approximates the operator
by $2m\sqrt{-\Box}$ (hence the weight is $(1+w)/2\simeq 1/2$
corresponding to $\,\ln\sqrt{-\Box}=(1/2)\ln(-\Box)\,$).

By decomposing the nonlocal fractions of (\ref{efacpert}) into the
sum of simple fractions and using the weighted proper time
representations (\ref{proptimerep1})-(\ref{proptimerep3}) and their
derivatives with respect to mass parameters $m_\pm$ we obtain the following expression
    \begin{eqnarray}
    \frac1{(M^2-\Box)^{k/2}}\,
    \frac1{(-\Box+2m\sqrt{M^2-\Box}\,)^l}=
    \int\limits_0^\infty\frac{ds}s\,
    e^{-s(M^2-\Box)}\,w_{kl}(s,m,M),     \label{w1}
    \end{eqnarray}
with some weight function $w_{kl}(s,m,M)$.\footnote{Alternatively
this weight function can of course be obtained as a Mellin transform
of the function of $\Box$ in the left hand side, but this simple
fraction decomposition method gives a more regular and systematic
way to achieve the needed goal.}

Using (\ref{zeroefac}) and (\ref{w1}) we finally obtain for the
perturbative expansion (\ref{efacpert})
    \begin{eqnarray}
    &&\frac12\;{\rm Tr}
    \ln \mbox{\boldmath$F$}^{\,\rm brane}=
    -\frac12\,
    \int\limits_0^\infty\frac{ds}s\,
    e^{-sM^2}\,\frac{w(-m_+\sqrt{s})+w(-m_-\sqrt{s})}2\;
    {\rm Tr}\,e^{\,s\,\Box}\nonumber\\
    &&\qquad\qquad\qquad\qquad+\sum_{k\geq 0,\,l\geq 1}\;
    \int\limits_0^\infty\frac{ds}s\,
    e^{-s\,M^2}\,w_{kl}(s,m,M)\;{\rm Tr}\Big[
    W_{kl}(\nabla)\,e^{\,s\,\Box}\Big].         \label{finalresult}
    \end{eqnarray}
This formally solves the problem of constructing the
Schwinger-DeWitt expansion for the brane effective action, because
as it was expected all the remaining calculations reduce to the
conventional calculation of the coincidence limits of the
Schwinger-DeWitt coefficients and their covariant derivatives in
    \begin{eqnarray}
    {\rm Tr}\Big[
    W_{kl}(\nabla)\,e^{\,s\,\Box}\Big]=\frac1{(4\pi s)^{d/2}}\int
    d^dx\,g^{1/2}\sum_{n=0}^\infty\,s^n\,
    {\rm tr}\,W_{kl}\Big(\nabla^x_\mu-\frac{\sigma_\mu(x,x')}{2s}\Big)\,
    \hat a_n(x,x')\,\Big|_{\,x'=x}.
    \end{eqnarray}
Remember that every $W_{kl}(\nabla)$ is a finite order covariant
differential operator with the coefficients built of the powers of
the bulk curvature, extrinsic curvature of the brane and their
covariant derivatives. Here lengthening of the derivatives in
$W_{kl}(\nabla)$ originates from commuting them with the the
exponential factor $\exp(-\sigma(x,x')/2s)$ contained in the kernel
of $\exp(s\Box)$, $\sigma_\mu(x,x')\equiv \nabla_\mu^x\sigma(x,x')$.
This of course brings to life world function coincidence limits
$\nabla_{\mu_1}...\nabla_{\mu_p}\sigma(x,x')|_{x'=x}$ also easily
calculable by the DeWitt recurrence procedure
\cite{DeWitt,PhysRep}.

It is important that the expansion (\ref{finalresult}) is efficient
for the purpose of obtaining the asymptotic $1/M$-expansion. Indeed,
in view of the weight function asymptotics (\ref{wasymp}) the
$w(-m_-\sqrt{s})$-parts of the overall $w_{kl}(s,m,M)$ (cf.
Eq.(\ref{wpm})) with $m_-<0$ are suppressed at $s\to\infty$ by
$e^{-sM^2}$ and, therefore, generate after the integration over $s$
the needed $1/M^2$-series. In the $w(-m_+\sqrt{s})$-parts the
integrand behaves as $e^{-s(M^2-m_+^2)}$, and generates the
$1/(M^2-m_+^2)\sim 1/2mM$-series also appropriate for the
$M\to\infty$ limit, though converging slower than the $1/M^2$ one.
This is the expansion in inverse squared masses of the real particle
associated with the pole (\ref{realparticle}). Unfortunately, the
powers of $1/M$ are accompanied by those of $1/m$, which comprises
in the DGP model the problem of low strong-coupling scale
\cite{scale} for small DGP crossover scale $m$.

\section{Conclusions}
  \hspace{\parindent}
This is obvious that the Schwinger-DeWitt technique in brane models
is much more complicated than in models without spacetime
boundaries. It does not reduce to a simple bookkeeping of local
surface terms like the one for simple boundary conditions reviewed
in \cite{Vassilevich}. Nevertheless it looks complete and
self-contained, because it provides in a systematic way a manifestly
covariant calculational procedure for a wide class of boundary
conditions including tangential derivatives (in fact of any order).
On the other hand, the calculational strategy of the above type is
thus far nothing but a set of blueprints for the Schwinger-DeWitt
technique in brane models, because there is still a large set of
issues and possible generalizations to be resolved in concrete
problems.

One important generalization is a physically most interesting limit
of the vanishing bulk mass $M^2$. Local curvature expansion is
perfect and nonsingular for nonvanishing $M^2$ and applicable in the
range of curvatures and magnitudes of spacetime derivatives
$(R,\,K^2,\nabla K)\ll M^2$, $\nabla\nabla R\ll M^4$, etc. However,
for $M^2\to 0$ it obviously breaks down, because the proper time
integrals start diverging at the upper limit. These infrared
divergences can be avoided by a nonlocal curvature expansion of the
heat kernel of \cite{CPT}. Up to the cubic order in curvatures this
expansion explicitly exists for ${\rm Tr}\,e^{s\Box}$ \cite{CPT3},
but for the structure involving a local differential operator ${\rm
Tr}\,W(\nabla) e^{s\Box}$ it still has to be developed.

Another important generalization is the extension of these
calculations to the cases when already the lowest order
approximation involves a curved spacetime background (i.e. dS or
AdS bulk geometry, deSitter rather than flat brane, etc.). The
success of the above technique is obviously based on the exact
knowledge of the $y$-dependence in the lowest order Green's function
in the bulk and the possibility to perform exactly (or
asymptotically for large $M^2$) the integration over $y$. All these
generalizations and open issues are currently under study.

To summarize, we developed a new scheme of calculating quantum
effective action for the braneworld DGP-type system in curved
spacetime. This scheme gives a systematic curvature expansion by
means of a manifestly covariant technique. Combined with the method
of fixing the background covariant gauge for diffeomorphism
invariance developed in \cite{gospel,qeastbg} this gives the
universal background field method of the Schwinger-DeWitt type in
gravitational brane systems.

\section*{Acknowledgements}
  \hspace{\parindent}
The work of A.B. was supported by the Russian Foundation for Basic
Research under the grant No 08-01-00737. The work of D.N. was partly
supported by the RFBR grant No 08-02-00725. D.N. is also grateful to
the Center of Science and Education of Lebedev Institute and Russian
Science Support Foundation. This work was also supported by the LSS
grant No 1615.2008.2.


\end{document}